# A maximum principle for the stochastic differential equations with multiplicative noise

Dietrich Ryter

ryterdm@gawnet.ch

Midartweg 3   CH-4500 Solothurn  Switzerland

Phone   +4132 621 13 07

Agreement of the probability current with the resolving paths requires a simplified forward equation for the (unique) Ito paths. Their increments are the most probable rather than expected ones, in accordance with an existing extremum principle. The latter is also generalized.

# I. Introduction

The textbook theory of the SDEs [1-7] is not self-consistent when the influence of the noise sources depends on the random state ("multiplicative noise"): (A) Solutions of the forward or Fokker-Planck equations (FPE) remain oblique for small time steps, while the path increments are Gaussian distributed. (B) The probability current disagrees with the path increments. The problem (A) was recently treated by the heuristic request that the density by the FPE and the Gaussian of the increments have their maximum at the same place [8]. The result was a new FPE for the unique Ito paths, now with most probable rather than expected increments. The present paper is focused on the problem (B). It derives the main results of [8] in a simpler way, and it mainly confirms them by a well-known extremum principle [6] which exhibits the role of the most probable displacements; this becomes crucial with the oblique solutions of the FPE. That principle even applies without the existence of a steady state (assumed for the quasipotential).

It turns out that the most probable displacements coincide with the noiseless motion, which is further shown to go to the maximum point of a steady density, if any. Time-dependent solutions of the FPE only describe the paths when the noise is so weak that the most probable and the expected values coincide. Steady densities may nevertheless exist with any noise level, under a specified condition.

The important results are covariant (i.e. compatible with any change of the variables), in particular the equilibrium densities and the prevailing time-dependent results. This is important in physics. The new FPE for the Ito paths may further be relevant in filtering and in stochastic control.

# II. Background

The continuous Markov process $\vec{X}(t)$ is assumed to fulfill the stochastic differential

equation (SDE)

$$dX^i = a^i(\vec{X})\,dt + b^{ik}(\vec{X})\,dW_k \quad \text{or} \quad d\vec{X} = \vec{a}(\vec{X})\,dt + \underline{B}(\vec{X})\,d\vec{W} \tag{2.1}$$

with smooth functions $a^i(\vec{x})$, $b^{ik}(\vec{x})$. The drift $\vec{a}$ is supposed to be independent of the noise. As usual, (2.1) denotes an integral equation, with the "integration sense" specified by $\alpha$ ($0 \leq \alpha \leq 1$; "Ito" for $\alpha = 0$, "Stratonovich" for $\alpha = 1/2$ and "anti-Ito" for $\alpha = 1$). The Wiener processes $W_k(t)$ are Gaussian distributed, with $<W_k(t) - W_k(0)> = 0$ and $<[W_k(t) - W_k(0)]^2> = t$; they are independent of each other.

The existing expression for the increments, with given $\vec{X}(t) = \vec{x}$ and $dt \geq 0$, is

$$\vec{X}(t+dt) - \vec{x} = \vec{a}(\vec{x})\,dt + \underline{B}(\vec{x})\,d\vec{W} + \alpha\,\vec{a}_{Sp}(\vec{x})\,dt + o(dt) \quad, \tag{2.2}$$

see [1-7,9], where $d\vec{W} := \vec{W}(t+dt) - \vec{W}(t)$, and with the "spurious" drift

$$a^i_{Sp}(\vec{x}) := b^{ij}{}_{,k}(\vec{x})\,b^{kj}(\vec{x}) = (\underline{B}_{,k}\underline{B}^T)^{ik} \,. \tag{2.3}$$

Note that (2.2) is Gaussian distributed, since the coefficients are taken at the initial $\vec{x}$. An essential role is played by the "diffusion matrix"

$$\underline{D}(\vec{x}) := \underline{B}(\vec{x})\,\underline{B}^T(\vec{x}) \,. \tag{2.4}$$

In [8] it was shown that $\vec{a}_{Sp}$ is typically given by the $\vec{x}$-dependence of $\underline{D}(\vec{x})$:

$$a^i_{Sp} = D^{ik}{}_{,k}/2 \,. \tag{2.5}$$

According to the standard literature the density $w(\vec{x},t)$ of $\vec{X}(t)$ (i.e. of the leading points of the random paths) is determined by the "forward equation"

$$w_{,t} = [-(a^i + \alpha\,a^i_{Sp})w + (1/2)(D^{ik}w)_{,k}]_{,i} \tag{2.6}$$

which is of the Fokker-Planck type. By $(D^{ik}w)_{,k} = D^{ik}{}_{,k}w + D^{ik}w_{,k}$ and (2.5) it can be written as

$$w_{,t} = \nabla \cdot \{-[\vec{a} + (\alpha-1)\vec{a}_{Sp}]w + \underline{D}\nabla w/2\} \,. \tag{2.7}$$

With the probability current

$$\vec{J}(\vec{x},t) := [\vec{a} + (\alpha - 1)\vec{a}_{Sp}]w - \underline{D}\nabla w/2 \qquad (2.8)$$

it becomes the continuity equation $w_{,t} + \nabla \cdot \vec{J} = 0$.

The "propagator" is defined as the solution of (2.7) with an initial deltafunction, asymptotically for small time steps.

### III. The path increments and the current

Consider a (moving or steady) smooth maximum of $w(\vec{x},t)$. There the probability current (2.8) reduces to $\vec{J}(\vec{x},t) = [\vec{a} + (\alpha - 1)\vec{a}_{Sp}]w$. The path increment (without the Gaussian deviations of mean zero) is given by $(\vec{a} + \alpha \vec{a}_{Sp})dt$. It must agree with the current, and this implies that

$$\vec{a} + (\alpha - 1)\vec{a}_{Sp} = \vec{a} + \alpha \vec{a}_{Sp} . \qquad (3.1)$$

By $\vec{a}_{Sp} \neq \vec{0}$, see (2.5), this is a contradiction for each $\alpha$. The textbook relation between SDE and FPE can therefore not hold when $\vec{a}_{Sp} \neq \vec{0}$. The following arguments will show that the drift in the FPE is not really determined by the *expected* path increment. For an alternative, consider the FPE with $\alpha = 1$, i.e.

$$w_{,t} = \nabla \cdot (-\vec{a} w + \underline{D} \nabla w/2) . \qquad (3.2)$$

According to [8] its propagator has a maximum moving by $\vec{a}\,dt$, while the asymmetric tails contribute $\vec{a}_{Sp}\,dt$ to the mean shift. The Ito increment (without $\underline{B}(\vec{x})d\vec{W}$) is therefore the *most probable* one, and *it agrees with the current $\vec{a}w$*. In other words: the random paths belonging to (3.2) are the Ito ones, but with most probable rather than expected increments. The last statement is confirmed by an existing extremum principle, see below.

## IV. The principle of the maximum probability

The evolution of a system is determined by the principle of the least action [6]. This will briefly be restated in a simple way. An attracting $\vec{a}(\vec{x})$ is supposed to admit a steady density $w(\vec{x},\infty)$. With a scaling factor $\varepsilon > 0$ for $\underline{D}$ and $\vec{a}_{Sp}$ (to exhibit the weak noise asymptotics, otherwise $\varepsilon = 1$) $w(\vec{x},\infty)$ is conveniently described by the "quasipotential" $\phi(\vec{x})$, defined by $w(\vec{x},\infty) := N(\varepsilon)\exp[-\phi(\vec{x})/\varepsilon]$. Inserting this into (2.7) yields

$$\nabla\phi \cdot (\vec{a} + \underline{D}\nabla\phi/2) - \varepsilon [\nabla \cdot (\vec{a} + \underline{D}\nabla\phi/2) + (1-\alpha)(\vec{a}_{Sp} \cdot \nabla\phi - \varepsilon \nabla \cdot \vec{a}_{Sp})] = 0. \quad (4.1)$$

The most important information is contained in the "eikonal equation"

$$\nabla\phi \cdot (\vec{a} + \underline{D}\nabla\phi/2) = 0 \quad (4.2)$$

which does not involve $\varepsilon$, $\alpha$ and $\vec{a}_{Sp}$. As a first order equation it can be solved by characteristics. These obey a Hamiltonian system of equations, where $\phi(\vec{x})$ is the action function (see e.g.[9], which also involves a simpler computation of $\phi$ in two dimensions). Actual motions minimize $\phi$ and therefore maximize $w$. This is crucial for the non-Gaussian propagators, where it makes the decision between the most probable and the expected evolution.

The arguments of the preceding Chapter III are not restricted to systems with a steady state. A new "principle of the most probable evolution" is therefore more general.

## V. The most probable path

The most probable continuation from a starting point $\vec{x}_0$ follows the drift $\vec{a}$ and is thus given by $\dot{\vec{x}}_\ell = \vec{a}(\vec{x}_\ell)$ with $\vec{x}_\ell(0) = \vec{x}_0$. The evolution of $w(\vec{x},t)$ along $\vec{x}_\ell(t)$ (starting with a deltafunction at $\vec{x}_0$) will be shown in the Appendix; the second derivatives on $\vec{x}_\ell(t)$ obey a Riccati equation. If (3.2) has a steady solution, $\vec{x}_\ell(t)$ tends to the location of its maximum. This can be inferred from (4.2) : if $\vec{a}(\vec{x})$ has an attracting point, $\vec{a}$ vanishes there, whence $\nabla \phi = \vec{0}$ ($\phi$ is minimum). The solution of the Riccati equation tends to the local quadratic solution of (4.2). On a more general compact attractor of $\vec{a}(\vec{x})$ $\phi$ is constant by the Lyapunov property of $\phi(\vec{x})$; note that $\vec{a} \cdot \nabla \phi = -(\nabla \phi)^T \underline{D} \nabla \phi / 2 < 0$ wherever $\nabla \phi \neq \vec{0}$. Knowledge of $\vec{x}_\ell(t)$ and of $w(\vec{x},t)$ along it is the prevailing dynamical information in many applications.

## VI. Discussion

6.1 *Restrictions on the densities*

Recall that the propagator involves $\vec{a}_{Sp}$, in contrast to the path increments. This means that time-dependent densities only describe the paths when $\vec{a}_{Sp}$ can be neglected, i.e. for weak noise.

The above argument does not apply for steady densities. They hold for each noise level when the neglected term in (4.1) really vanishes, i.e. when

$$\nabla \cdot (\vec{a} + \underline{D} \nabla \phi / 2) = 0 \ . \tag{6.1}$$

This expresses the local (detailed) balance between dissipation and diffusion. When it does not hold, a time-dependence persists for $t \to \infty$.

6.2 *Covariance*

Any displacement in $\vec{x}$-space is a contravariant vector, by definition. This implies the covariance of the random paths. While $\vec{a}(\vec{x})$ is contravariant [5,8], $\vec{a}_{Sp}(\vec{x})$ is *not* a tensor, since by (2.5) it vanishes in the variables $\vec{z}(\vec{x})$ that yield a constant $\underline{D}$ [10]. This immediately singles out the Ito paths (the rows of $\underline{B}$ must be contravariant, as $\vec{W}(t)$ is not transformed). Note that $\vec{a}_{Sp}$ does not explicitly occur in (3.2), but in its propagator. Time-dependent densities are thus only covariant for weak noise, where $\vec{a}_{Sp}$ in the propagator is neglected against $\vec{a}$. Strictly steady densities, on the contrary, are covariant. The "eikonal equation" (4.2) is covariant for a scalar $\phi$, since $\underline{D}$ is a twice contravariant tensor [5]. The extra condition (6.1) means that $\vec{a} + \underline{D}\nabla\phi/2$ is source-free (solenoidal), and a contravariant field preserves that property in other variables.

Covariance thus holds under the restrictions of the above § 6.1 .

**Appendix:** *The Gaussian density along the noiseless path*

The peak of a density starting with $\delta(\vec{x} - \vec{x}_0)$ moves along $\vec{x}_\ell(t)$. The aim is to evaluate the peak value and the Gaussian approximation along that path.

The time-derivative of the peak value $w[\vec{x}_\ell(t)]$ is $\dot{w} = w_{,t} + \nabla w \cdot \dot{\vec{x}}_\ell$ . By $\nabla w = \vec{0}$ on $\vec{x}_\ell(t)$ it reduces to $\dot{w} = \nabla \cdot (-\vec{a}w + \underline{D}\nabla w/2)$. It is advantageous to introduce the time-dependent quasipotential $\Phi(\vec{x},t) := -\log w(\vec{x},t)$, by which $\nabla w = -w\nabla\Phi$, so that

$$\dot{w} = -\nabla \cdot [w(\vec{a} + \underline{D}\nabla\Phi)] = -w\nabla \cdot (\vec{a} + \underline{D}\nabla\Phi) = -w[\nabla \cdot \vec{a} + tr(\underline{D}\,\underline{S})] \ , \tag{A.1}$$

where $\underline{S}(t)$ is the matrix of the second derivatives of $\Phi$. This allows to compute $w[\vec{x}_\ell(t)]$ when $\underline{S}$ is known on $\vec{x}_\ell(t)$. The change of $\underline{S}(t)$ in $dt$ is determined by the Gaussian of the Ito increment. Using the matrix $\underline{M}$ with the elements $(a)^i_{,k}$ one can derive from [11] that $\underline{S}[\vec{x}_\ell(t+dt)] = \underline{S} + (\underline{S}\,\underline{M} + \underline{M}^T\underline{S} + 2\underline{S}\,\underline{D}\,\underline{S})dt$, i.e. the Riccati equation

$$\underline{\dot{S}} = \underline{S}\,\underline{M} + \underline{M}^T\underline{S} + 2\underline{S}\,\underline{D}\,\underline{S} \ , \tag{A.2}$$

and multiplying from both sides by the inverse $\underline{S}^{-1} := \underline{Q}$ results in the linear equation

$$\underline{\dot{Q}} = \underline{M}\,\underline{Q} + \underline{Q}\,\underline{M}^T + 2\underline{D} \ . \tag{A.3}$$

The initial $\underline{Q}$ is zero, and the final $\underline{\dot{Q}} = \underline{0} = \underline{\dot{S}}$ yields the quadratic expansion of the steady quasipotential at a point attractor of $\vec{a}(\vec{x})$, see e.g. [12,13]. The starting values of $\underline{S}$ and $w[\vec{x}_\ell(t)]$ can be obtained from the Gaussian solution after $dt$.